\documentclass[12pt]{iopart}

\usepackage{iopams}
\usepackage{amsfonts}
\usepackage{amsbsy} 
\usepackage{epsfig}
\usepackage{bbm}
\usepackage{latexsym}

\def\ben{\begin{equation}}
\def\een{\end{equation}}
\def\bea{\begin{eqnarray}}

\def\eea{\end{eqnarray}}

\def\const{\rm constant}

\def\bR{\mathbb R}

\def\bZ{\mathbb Z}
\def\b1{e^0}

\def\im{{\rm i}}
\def\f{{\rm f}}
\def\rme{{\rm e}}
\def\rmd{{\rm d}}
\def\cV{{\cal V}}
\newcommand{\arx}[1]{(#1)}

\begin{document}

\begin{flushright} Class. Quantum Grav. {\bf 28} (2011) 125014 \hfill arXiv:1011.4290\\ AEI-2010-162 \\ DAMTP-2010-91\end{flushright}

\title{Two-point functions in (loop) quantum cosmology}

\author{Gianluca Calcagni$^1$, Steffen Gielen$^{1,2}$ and Daniele Oriti$^1$}
\address{$^1$ Max Planck Institute for Gravitational Physics (Albert Einstein Institute),
Am M\"uhlenberg 1, D-14476 Golm, Germany}
\address{$^2$ DAMTP, Centre for Mathematical Sciences, Wilberforce Road, Cambridge CB3 0WA, UK} 
\ead{calcagni@aei.mpg.de{\rm ,} gielen@aei.mpg.de {\rm and} doriti@aei.mpg.de}

\date{November 18, 2010}

\begin{abstract}
The path-integral formulation of quantum cosmology with a massless scalar field as a sum-over-histories of volume transitions is discussed, with particular but non-exclusive reference to loop quantum cosmology. Exploiting the analogy with the relativistic particle, we give a complete overview of the possible two-point functions, pointing out the choices involved in their definitions, deriving their vertex expansions and the composition laws they satisfy.
We clarify the origin and relations of different quantities previously defined in the literature, in particular the tie between definitions using a group averaging procedure and those in a deparametrized framework. Finally, we draw some conclusions about the physics of a single quantum universe (where there exist superselection rules on positive- and negative-frequency sectors and different choices of inner product are physically equivalent) and multiverse field theories where the role of these sectors and the inner product are reinterpreted.
\end{abstract}

\pacs{98.80.Cq, 04.60.Pp}

\maketitle


\section{Introduction}

One of the major problems of modern theoretical physics is the resolution of singularities in the structure of classical spacetime. In particular, the big bang problem in General Relativity has been inspiring a steady trend of research aiming at the removal of the initial singularity via quantum effects, i.e., by replacing classical General Relativity with a quantum theory of spacetime and gravity. The formal development of a background-independent theory of quantum gravity is no easy task and some physical insight is gained by looking at a mini-superspace toy model where the gravitational degrees of freedom are reduced to a minimum. On this Friedmann--Robertson--Walker (FRW) background, one can write the gravity and matter Hamiltonian and promote it to an operator on a Hilbert space of states associated with the probability for the universe to be in a given phase-space configuration. The ensuing quantum theory can then be studied by exact or perturbative techniques and the fate of the big bang singularity examined. While traditional quantum cosmology corresponded to a symmetry-reduced version of the canonical quantum gravity in ADM variables \cite{admrefs}, a new scheme for quantum cosmology, called loop quantum cosmology (LQC) \cite{LivRev}, has been developed drawing inspiration from loop quantum gravity (LQG), an approach to the canonical quantization of gravity based on connection variables \cite{thomas,carlo}. While the traditional Wheeler--DeWitt quantization generally fails to solve the problem, loop quantum cosmology does avoid the singularity by replacing it with a bounce at a finite curvature and matter density. Unfortunately, the robustness of the results of LQC depend on the yet unclear connection between the model and the full theory of loop quantum gravity, itself still under active development and far from being complete. In particular, concerning the full theory, most recent activities towards a complete definition of the quantum dynamics have focused on a covariant or path integral version of the theory, going under the name of spin foam models \cite{alex,danielethesis}. Just as LQG states are graphs labelled by group elements or group representations, spin foam histories are 2-complexes labelled by the same type of data. Recently, a path-integral formulation has been proposed also for LQC \cite{ach,achlong}. The main virtues of this reformulation are the following: (a) obviously, this alternative definition of the quantum dynamics could be advantageous from the purely technical point of view as well as offer new conceptual insights; (b) it recasts the dynamics of LQC in a form that formally resembles the spin foam formulation of the full theory; in particular, it is expressed in terms of a `vertex expansion' (see below and \cite{ach, achlong}) and could be used as testing ground for techniques later to be applied to spin foam models; (c) it could exemplify issues and conceptual points that one would eventually have to address in the spin foam formulation of quantum gravity and suggest solutions to the same. On this last point, several interesting questions and suggestions already came out of the work in \cite{ach,achlong} and from the follow-up work (e.g.~\cite{hrvw}), concerning the possibility of defining different types of two-point functions in LQC, which maybe encode different causal properties and dynamical information. In fact, this point had been raised earlier in the quantum gravity literature \cite{Tei82, halliwellhartle} and later in the spin foam literature \cite{causalBC,feynman,causal3dmatter}.

The covariant construction of two-point functions is particularly simple because of the close analogy between flat cosmology with a massless scalar field and the relativistic, massive point particle in quantum mechanics \cite{Tei82,HaKu}. In fact, not only do both systems have the same number of degrees of freedom, but they are also formally identical.\footnote{The analogy with the relativistic particle not only holds for the symmetry-reduced theory but also in full general relativity \cite{Tei82, halliwellhartle,HaKu,BSW}.} However, because of the very different physics they entail, some aspects and consequences of the path integral formalism have not been fully elucidated, causing some results in the literature to appear unexpected and somewhat unclear. The main purpose of this paper is to further investigate the structure of the mini-superspace path integral formulation, and catalogue the whole zoo of two-point functions\footnote{Unlike some other authors in the physics literature, we reserve the term `Green's function' for propagators, i.e.~two-point functions that solve the equation of motion with a delta source.} that can be constructed for this system. In particular, we will analyze the mutual relations between (a) different representations of the physical Hilbert space, (b) different frequency sectors, (c) two-point functions and the composition laws governing the evolution of the quantum universe, and (d) different time formalisms. Thanks to a complete comparison with the single particle, the combination of all these aspects in a self-consistent framework will allow us to acquire a precise dictionary and some novel physical insights.

In particular, and summarizing briefly our results, the single universe picture is the analogue of the single massive particle in quantum mechanics, where the path integral is the one-dimensional sum over histories of an individual quantum object. A two-point function governs the evolution of the universe from one volume/matter configuration to another. One can choose between different but physically equivalent canonical inner products, both obtainable from a group averaging procedure, depending on whether one uses a relativistic or non-relativistic representation of the canonical states. The first choice (causal two-point function) is non-positive-definite because positive- and negative-frequency sectors contribute with opposite signs; however, these sectors are superselected by a complete set of Dirac observables and one can take either. Picking positive-frequency modes only (thus obtaining the positive Wightman function), the relativistic inner product is well-defined. The non-relativistic inner product (Newton--Wigner function) is always well-defined. Both restricted two-point functions satisfy an appropriate composition law. In the relativistic representation, one could also define an inner product that is positive definite and includes both sectors, the Hadamard function, but this fails to satisfy a nice composition law. Moreover, one can define proper transition amplitudes (true Green functions for the Hamiltonian constraint) both in the full Hilbert space (Feynman propagator) or for a single frequency sector (retarded/advanced propagator), and impose causality, by restricting the one-parameter integral over proper time (Schwinger representation) from the full real line (Hadamard function) to the positive half-line (Feynman propagator). All such transition amplitudes satisfy the appropriate composition law.
Also in the case of (loop) quantum cosmology, in the presence of a massless scalar field, as we will show, there exist physically equivalent representations of the two-point functions, as well as an identical catalogue of them, obtained in roughly the same manner, and depending on the same choices. 
These results hopefully settle some recent debate about the origin of different definitions for the inner product in cosmology, and their mutual relations. Such issues are also discussed from the perspective of building a `third quantized' or multiverse theory, where the role of different sectors and the choice of inner product are drastically reinterpreted. We stress that our findings hold independently of the form of the gravitational part of the Hamiltonian constraint: while they apply rigorously to loop quantum cosmology where the gravitational configuration variable $\nu$ defined below takes discrete values, they can be extended at least formally to the standard Wheeler--DeWitt equation, upon the replacement of summations over $\nu$ by integrations.


\section{Basics of (loop) quantum cosmology and particle analogy}

We will concentrate on the simplest cosmological model of a spatially flat, isotropic universe with a massless scalar field as matter. For a canonical analysis of this model, one first fixes a fiducial three-dimensional cell of comoving volume $\cV_0<\infty$ to avoid divergences associated with infinite spatial volume, together with a flat metric ${}^0 q_{ab}$ which may be taken to be $\delta_{ab}$ in Cartesian coordinates.\footnote[1]{Here and in the following, spatial tangent indices are denoted by $a,b,c$ and internal $\mathfrak{su}(2)$ indices are denoted by $i,j,k$.} This model admits a precise canonical analysis both in the Wheeler--DeWitt quantization and in LQC; the difference between the two approaches lies in the choice of canonical variables and in the operator reordering in the Hamiltonian constraint, in turn leading to different quantum dynamics.\footnote{Strictly speaking, there is also an intermediate step before quantization, making also the classical dynamics different. In fact, after rewriting the classical Hamiltonian constraint in terms of holonomy plaquettes and positive powers of the triad, one ends up with an expression where the limit of zero area for the plaquettes is taken. A non-trivial {\it ad hoc} step, inspired by the LQG kinematics, is to regularize the Hamiltonian by imposing a non-zero lower bound for the holonomy areas, and \emph{then} quantize.}


\subsection{Kinematics}

Focussing on the gravitational variables first, after a choice of orthonormal frame $\{{}^0 e^a_i\}$ and dual $\{{}^0 e_a^i\}$ compatible with ${}^0 q_{ab}$ one may parametrize the Ashtekar--Barbero $\mathfrak{su}(2)$ connection $A_a^i$ and the densitized triad $E^a_i$ by
\ben
A_a^i = \frac{c}{\cV_0^{1/3}}\;{}^0 e_a^i\,,\qquad E^a_i=\frac{p \sqrt{\det \,{}^0 q}}{\cV_0^{2/3}}\;{}^0 e^a_i\,,
\een
where powers of $\cV_0$, the 3-volume of the fiducial cell as measured by ${}^0 q_{ab}$, have been introduced to make the coordinates $(c,p)$ invariant under rescalings of ${}^0 q_{ab}$. The coordinates are canonically conjugate variables,
\ben
\{c,p\}=\frac{8\pi G\gamma}{3}\,,
\een
where $\gamma$ is the Barbero--Immirzi parameter, and the only non-trivial dynamics of the dimensionally reduced system is encoded in the Hamiltonian constraint of first-order general relativity.

While in traditional mini-superspace (Wheeler--DeWitt) approaches to quantum cosmology one quantizes the system in the standard Schr\"odinger representation, where $\hat{c}$ and $\hat{p}$ are operators on a (kinematical) Hilbert space, in LQC one follows the kinematics of full loop quantum gravity, quantizing only holonomies of the connection instead of the connection itself. That is, instead of $\hat{c}$ and $\hat{p}$ one defines $\hat{p}$ and $\widehat{\exp({\rm i}\mu c)}$. Here, $\mu$ denotes a real parameter which may also be chosen as a function of $p$ by means of a suitable procedure. The precise form of $\mu$ will not be relevant for our analysis.

The kinematical Hilbert space $\mathcal{H}_{{\rm kin}}^{{\rm g}}$ is taken to be the space of square integrable functions on the Bohr compactification of the real line.\footnote{The algebra of continuous functions on this space is equivalent to the almost periodic functions on $\bR$ as introduced in \cite{bohr}.} One usually works in a basis where $\hat{p}$ is diagonal, with orthonomality relation
\ben\nonumber
\langle p | p' \rangle = \delta_{p,p'}\,.
\een
Since this is a Kronecker delta rather than a distribution, one is dealing with a non-separable Hilbert space. If $\mu$ is taken to be a non-trivial function of $p$, the action of the holonomy operator $\widehat{\exp({\rm i}\mu c)}$ takes a rather complicated form in this representation, so that its
commutator with $p$ is not simply a multiple of the identity. It is usually more convenient \cite{acs} to pass to a basis $\{|\nu\rangle\}$ of eigenstates of the volume operator $\hat{V}$ measuring the kinematical volume of the fiducial cell, $V=|p|^{3/2}$,
\ben
\hat{V}|\nu\rangle = 2\pi\gamma \, |\nu|\,l_{{\rm Pl}}^2 |\nu\rangle\,.
\een
Note that $\nu$ has dimensions of length. One can normalize the states $\{|\nu\rangle\}$ so that one still has an orthonormal basis,
\ben
\langle \nu | \nu' \rangle = \delta_{\nu,\nu'}\,.
\een
The basic operators are now $\hat\nu$, which acts by multiplication, and $\widehat{\exp(\im\lambda b)}$, where $b$ is conjugate to $\nu$ and $\lambda={\rm const}$, which acts as a shift in $\nu$. These satisfy the standard Heisenberg algebra. For the matter sector, one chooses a standard Schr\"odinger quantization with a natural representation of the Hilbert space $\mathcal{H}_{{\rm kin}}^{\phi}$, the space of square integrable functions on $\bR$, on which $\hat{\phi}$ acts by multiplication and $\hat{p_{\phi}}$ by derivation, and with an orthonormal basis given by 
\ben\label{phino}
\langle \phi | \phi' \rangle = 2\pi\delta(\phi-\phi')\,,
\een
where the $2\pi$ factor is for later convenience. The Hilbert space of the coupled system is then just the tensor product $\mathcal{H}_{{\rm kin}}^{{\rm g}} \otimes \mathcal{H}_{{\rm kin}}^{\phi}$.


\subsection{Dynamics}

To obtain the space of physical states, one is now left with solving the Hamiltonian constraint which can be written in the form
\ben
\hat{C}\,\Psi(\nu,\phi) \equiv -\left(\Theta+\partial_{\phi}^2\right)\,\Psi(\nu,\phi)\,,
\label{hamconst}
\een
where $\Psi$ is a wavefunction on configuration space and $\Theta$ only acts on $\mathcal{H}_{{\rm kin}}^{{\rm g}}$. In LQC, $\Theta$ is a difference operator which typically takes the form (e.g.~\cite{ach})
\ben\label{thet}
\Theta\Psi(\nu,\phi) := A(\nu)\Psi(\nu+4l_0,\phi)+B(\nu)\Psi(\nu,\phi)+C(\nu)\Psi(\nu-4l_0,\phi)\,,
\een
where $A,B,C$ are functions (dependent on the details of the quantization scheme) and $l_0$ is an elementary length unit, usually defined by the square root of the area gap encountered in loop quantum gravity (i.e.~the Planck length up to a numerical factor). In Wheeler--DeWitt theory, $\Theta$ is a differential operator acting on $L^2(\bR)$. The form of the operator $\Theta$ in either theory will not be important in the following, except for the fact that in LQC there is an interval's worth of superselection sectors in $\mathcal{H}_{{\rm kin}}^{{\rm g}}$; namely, $\Theta$ preserves all subspaces spanned by $\{|\nu_0+4nl_0\rangle~|~n\in\bZ\}$ for some $\nu_0$. We may restrict to one of these subspaces, i.e.~assume that wavefunctions only have support on a discrete lattice which we take to be $4l_0\bZ$ (see the comments in \cite{KLP} if one wants to keep the most interesting point $\nu=0$ in this lattice for a generic gauge choice). This restriction picks out a separable subspace of the originally non-separable Hilbert space. Our analysis will naturally apply to any $\Theta$ with this property in the setting of LQC, and to any differential operator in Wheeler--DeWitt theory in the formal sense that we will explain below. Our analysis will make no prior assumptions about the exact values of $\nu$ included in the lattice. We will sum over all of these in a given, unspecified range. Then, the results will be valid also for models of LQC where the range of $\nu$ is a subset of the integer numbers, for instance the positive semiaxis as in \cite{MMO1}.

Moreover, the space of states solving equation (\ref{hamconst}) is reducible in terms of two sectors of solutions satisfying, respectively, $(\hat{p}_\phi\,\mp\sqrt{\Theta})\Psi_{\pm}(\nu,\phi) =0$, resulting from the fact that $\hat{p}_\phi$ is a Dirac observable for the system governed by the above dynamics. This will allow us to perform an explicit splitting of the space of physical states into positive- and negative-frequency states.


\subsection{Particle analogy}

The form of the constraint (\ref{hamconst}) is that of a relativistic particle in 1+1 dimensions (with a discretized spatial coordinate in the case of LQC), where $\Theta$ plays the role of a second spatial derivative (up to a mass terms), and we think of a parametrized form of the particle action
\ben
S=\int\limits_{\tau'}^{\tau''}\rmd\tau\;\left[p_x \dot{x}+p_t\dot{t} + N(p_x^2-p_t^2+m^2)\right]\,,
\label{partact}
\een
which is invariant under reparametrizations $\tau\rightarrow f(\tau)$ of the worldline (here $\dot{}=\rmd/\rmd\tau$).

In the particle analogy, the scalar field $\phi$ plays the role of the the time coordinate $t$, and its conjugate momentum $p_\phi$ of the energy $E$, with the corresponding eigenstates related by Fourier transform. Correspondingly, the mentioned two sectors of solutions can be understood as the analogue of the positive- and negative-frequency sectors, respectively \cite{BlI} (see \cite{APS0} for a more recent discussion). Defining
\ben\label{fou}
|p_{\phi}\rangle \equiv \int  \frac{\rmd \phi}{2\pi}\, \sqrt{2|p_\phi|}\,\rme^{\im p_\phi \phi} \, |\phi\rangle\,, 
\een
one verifies
\ben
\hat p_\phi|p_\phi\rangle = p_\phi |p_\phi\rangle\,,\qquad p_\phi\in\bR\,,
\een
as well as the normalization
\ben\label{posidef}
\langle p_{\phi} | p_{\phi}' \rangle = 2|p_\phi|\,\delta(p_\phi-p_\phi')\,.
\een
Alternatively, one may remove the absolute value in the definition (\ref{fou}), to obtain the  indefinite normalization
\ben\label{negadef}
\langle p_{\phi} | p_{\phi}' \rangle = 2 p_\phi \,\delta(p_\phi-p_\phi')\,.
\een

For our purposes, the volume eigenstates $| \nu \rangle$ can equivalently be understood as playing the role of the spatial coordinates ${\bf x}$ (or of the spatial momenta ${\bf p}$, as our manipulations are performed always in the $\nu$ basis). We will make use of this analogy to extend what is known about two-point functions for a relativistic particle to the case at hand, in particular, their composition laws as analysed in \cite{halliortiz,HaT}. We note that the inclusion of a scalar field simplifies the formalism due to the existence of an internal `time' variable, but it is also important in order to give an operational meaning to a quantum system describing the universe as a whole, in particular in this case where gravity is approximated by a single degree of freedom.

As we will show in detail, entirely as in the particle case, the definition of different two-point functions is fully characterized by the following.
\begin{itemize} 
\item The canonical representation chosen for the physical quantum states in configuration space (more precisely, in the $\phi$ space). Typically, one has three choices. The first is a \emph{non-relativistic} representation\footnote{The name is inherited from the particle case, where it refers to the invariance properties of the states under $SO(1,1)$ transformations. In the LQC case, due to the discrete nature of the $\nu$ basis, the transformation properties under the same group are trickier at the quantum level. We maintain the nomenclature nevertheless.} in terms of states
\ben
| \nu , \phi\rangle_{\rm nr}\equiv | \nu , \phi ; +\rangle_{\rm nr}+| \nu , \phi ; -\rangle_{\rm nr}\,,
\een
where
\ben
| \nu , \phi ; \pm\rangle_{\rm nr} \, =\, \int \frac{\rmd p_\phi}{\sqrt{2|p_\phi|}}\,\rme^{\pm\im p_\phi \phi}\, \delta(p_\phi\,\mp\, \sqrt{\Theta})\,|\nu , p_\phi\rangle\,. \label{bothnon}
\een
These states are orthogonal at equal values of $\phi$ in the physical inner product and, choosing the positive-definite normalization (\ref{posidef}) and the suitable positive-definite resolution of the identity in momentum space, they satisfy the completeness relation
\ben\label{res1}
\mathbbm{1} \,=\, \sum_{\nu}\,\Big(  | \nu , \phi; +\rangle_{\rm nr}\,\langle \nu, \phi ; +|_{\rm nr} +  | \nu , \phi; -\rangle_{\rm nr}\,\langle \nu, \phi ; -|_{\rm nr}\Big)\,.
\een
This expression and the normalization factors in equation (\ref{bothnon}) depend on the details of the normalization of kinematical states in Fourier representation. One can always make all these definitions mutually consistent.

The second possibility is a \emph{relativistic} representation in terms of states
\ben
| \nu , \phi\rangle\equiv | \nu , \phi ; +\rangle+| \nu , \phi ; -\rangle\,,
\een
where
\ben
| \nu , \phi ; \pm \rangle\, =\, \int \frac{\rmd p_\phi}{2|p_\phi|}\,\rme^{\pm\im p_\phi \phi}\, \delta(p_\phi\,\mp\, \sqrt{\Theta})\,|\nu , p_\phi\rangle\,, \label{both}
\een
which are not orthogonal at fixed $\phi$. Choosing the indefinite normalization (\ref{negadef}) and the corresponding indefinite resolution of the identity in momentum space, these states satisfy the completeness relation 
\ben
\mathbbm{1} \,=\,\im \sum_{\nu}\, \left( | \nu , \phi; +\rangle\,\stackrel{\leftrightarrow}{\partial}_{\phi}\,\langle \nu, \phi ; + | \, +\,  | \nu , \phi; -\rangle\,\stackrel{\leftrightarrow}{\partial}_{\phi}\,\langle \nu, \phi ; - | \right) \,, \label{dec}
\een
where $\stackrel{\leftrightarrow}{\partial}=\stackrel{\rightarrow}{\partial}-\stackrel{\leftarrow}{\partial}$. It is immediate to see \cite{halliortiz} that the corresponding inner product, in the relativistic representation, is {\it not} positive definite (it is positive for the positive-frequency sector, and negative for the negative-frequency one). This would suggest that the corresponding quantum canonical theory is not well defined. A similar issue is present also in the full (formal) quantum geometrodynamics in Wheeler--DeWitt form \cite{BlI,APS0,kuchar}. One can obtain a positive definite inner product in the relativistic representation, and in the presence of both sectors of solutions, by choosing the positive-definite normalization (\ref{posidef}) instead of (\ref{negadef}), so that the decomposition (\ref{dec}) is replaced by
\ben
\mathbbm{1} \,=\, \im\sum_{\nu}\, \left( | \nu , \phi; +\rangle\,\stackrel{\leftrightarrow}{\partial}_{\phi}\,\langle \nu, \phi ; + | \, -\,  | \nu , \phi; -\rangle\,\stackrel{\leftrightarrow}{\partial}_{\phi}\,\langle \nu, \phi ; - | \right) \,.\label{dec2}
\een
However, we will show that such non-standard resolution of the identity affects the composition properties of the two-point function defining the physical inner product.

\item Whether one restricts consideration to either positive- or negative-frequency sectors of the Hilbert space, or works in the full Hilbert space. As we have just seen, the inclusion of both sectors has important consequences already at the canonical level. However, the presence of
a complete set of Dirac observables induces a superselection rule which allows one to pick only one type of modes. In particular, taking the positive-frequency sector the three resolutions of the identity (\ref{res1}), (\ref{dec}) and (\ref{dec2}) collapse into two positive-definite expressions, one nonrelativistic,
\ben
\mathbbm{1} \,=\, \sum_{\nu}\, | \nu , \phi; +\rangle_{\rm nr}\,\langle \nu, \phi ; + |_{\rm nr}\,,
\een
and one relativistic:
\ben
\mathbbm{1} \,=\, \im\sum_{\nu}\, | \nu , \phi; +\rangle\,\stackrel{\leftrightarrow}{\partial}_{\phi}\,\langle \nu, \phi ; +| \,.
\een
The two representations are different in many respects, but physically equivalent, as they are related by a unitary transformation \cite{BlI}.\footnote{This can be seen by interpreting the states in each sector as distributions over an auxiliary Hilbert space ${\cal S}$, and noticing that the definitions (\ref{bothnon}) and (\ref{both}) involve different spaces ${\cal S}$.}
\item The integration range in proper time (lapse) in the group averaging representation of the two-point function, or, equivalently, the class of histories summed over in the sum-over-histories formulation of the same. Integrals over the full real line give, generically, solutions of the constraint equation, and thus true inner products for the canonical theory; integrals of the positive semiaxis only give Green's functions (propagators) for the same constraint equation (i.e.~solutions of the constraint equation in the presence of a delta source).
\end{itemize}

We now proceed to define the various two-point functions for (loop) quantum cosmology, depending on the above choices, and to give, for each of them, their covariant (spin foam) representation in terms of sum-over-histories. This will also clarify their mutual relations, and the origin of their differences. Having done so, we proceed to derive the composition laws that the different two-point functions satisfy, and discuss their physical implications.


\section{Two-point functions as inner products in (loop) quantum cosmology}

Since one has a constrained system, the first quantity of interest is the physical inner product between states, which is usually defined, following Dirac \cite{dirac}, starting from generic (kinematical) states, labelled by unrelated $\nu$ and $p_\phi$, by inserting a projector onto those states annihilated by the constraints. Such an inner product defines a solution for the Hamiltonian constraint operator. It is the analogue of a transition amplitude for systems with an external time variable.


\subsection{Timeless and deparametrized frameworks}

In the cosmological model we are considering, the inner product between two states $|\nu_\im,\phi_\im\rangle$ and $|\nu_\f,\phi_\f\rangle$ can be defined rigorously
in the group averaging procedure (for a review, see \cite{groupav}). In the relativistic representation (\ref{both}) extended to both frequency sectors, the inner product is
\ben
G_{{\rm H}}(\nu_\f,\phi_\f;\nu_\im,\phi_\im)\equiv\int\limits_{-\infty}^{+\infty}\rmd\alpha\;\langle \nu_\f,\phi_\f|\rme^{\im\alpha \hat{C}}|\nu_\im,\phi_\im\rangle\,.
\label{hadamard}
\een
A slightly modified definition, obtained by working in the non-relativistic representation (\ref{bothnon}) also extended to both frequency sectors (once more, leading to a unitarily equivalent representation of Dirac observables), is used in \cite{achlong}, where an additional factor $2|p_{\phi}|$ is introduced in order to simplify calculations: 
\ben
G_{{\rm NW}}(\nu_\f,\phi_\f;\nu_\im,\phi_\im)\equiv\int\limits_{-\infty}^{+\infty}\rmd\alpha\;\langle \nu_\f,\phi_\f|\rme^{\im\alpha \hat{C}}2|\hat{p_{\phi}}||\nu_\im,\phi_\im\rangle\,.
\label{hadaach}
\een
A simple (but, in LQC, apparently unnoticed) fact is the following. For the relativistic particle, the analogue of (\ref{hadamard}) is the \emph{Hadamard function}, hence the suggestive notation $G_{{\rm H}}$. On the other hand, the analogue of (\ref{hadaach}), where a factor of the frequency is included, leads to a non-relativistic two-point function, namely the positive- or negative-frequency {\em Newton--Wigner function} if one restricts to positive or negative frequencies, or a sum of the two if one includes both. We will show that both expressions (\ref{hadamard}) and (\ref{hadaach}) admit exactly the same interpretation in quantum cosmology and obey the expected composition laws. In particular, equation (\ref{hadaach}) is the sum of positive- and negative-frequency Newton--Wigner functions.

The Hadamard function has a sum-over-histories representation \cite{halliortiz} obtained by fixing the reparametrization invariance of the action (\ref{partact}) in proper-time gauge $\dot{N}=0$:
\ben
G_{{\rm H}}(x''t'';x',t')=\int\limits_{-\infty}^{+\infty} \rmd T\;g(x'',t'';T|x',t';0)\,,
\label{hadasumoverh}
\een
where $g(x'',t'';T|x'',t';0)$ is a non-relativistic transition amplitude for the Hamiltonian $H=p_x^2-p_t^2+m^2$, in (proper) time $T$.

While this proper-time representation is analogous to what is done in full quantum gravity, in the cosmological setting one can also exploit the existence of a relational time $\phi$ to pass to a `deparametrized' formalism in which one takes the positive square root of (\ref{hamconst}) \cite{BlI},
\ben
-\im\partial_{\phi}\Psi(\nu,\phi)=\sqrt{\Theta}\Psi(\nu,\phi)\equiv H\Psi(\nu,\phi)\,,
\een
interprets the system as a non-relativistic particle in one dimension with external time $\phi$ (and Hilbert space $\mathcal{H}_{{\rm kin}}^{{\rm g}}$), and computes the transition amplitude
\ben
G_{{\rm NW}}^+(\nu_\f,\phi_\f;\nu_\im,\phi_\im)\equiv\langle \nu_\f|\rme^{\im H\Delta\phi}|\nu_\im\rangle\,,
\label{newtonwig}
\een
where we introduced the shorthand $\Delta\phi\equiv\phi_\f-\phi_\im$. Similarly, taking the negative square root of (\ref{hamconst}) one can define the negative-frequency amplitude
\ben
G_{{\rm NW}}^-(\nu_\f,\phi_\f;\nu_\im,\phi_\im)\equiv\langle \nu_\f|\rme^{-\im H\Delta\phi}|\nu_\im\rangle\,.
\label{newtonwig2}
\een
Obviously, this interpretation goes also behind the non-relativistic canonical representation given above.

For the relativistic particle, the non-relativistic functions (\ref{newtonwig}) and (\ref{newtonwig2}) are, again, the (positive- and negative-frequency) Newton--Wigner functions \cite{NW} (see also \cite{HaKu,Rui,ReS}), obtained from a sum-over-histories representation in canonical gauge $\phi=\tau$, where $\tau$ was the affine parameter in (\ref{partact}) \cite{halliortiz}. This is precisely what is done in the deparametrized formalism for (loop) quantum cosmology.

We now show that
\ben\label{rela}
G_{{\rm NW}}(\nu_\f,\phi_\f;\nu_\im,\phi_\im)= G_{{\rm NW}}^+(\nu_\f,\phi_\f;\nu_\im,\phi_\im)+ G_{{\rm NW}}^-(\nu_\f,\phi_\f;\nu_\im,\phi_\im)\,.
\een
The left-hand side is equation (\ref{hadaach}). Writing the integral in the parameter $\alpha$ as a delta distribution and using the formal relation
\ben\label{delta}
2|p_{\phi}|\delta(p_{\phi}^2-\Theta) = \delta(p_{\phi}-\sqrt{\Theta})+\delta(p_{\phi}+\sqrt{\Theta})\,,
\een
we get
\bea
\fl G_{{\rm NW}}(\nu_\f,\phi_\f;\nu_\im,\phi_\im) &=& 2\pi\langle \nu_\f,\phi_\f|2|\hat{p_{\phi}}|\delta(p_\phi^2-\Theta)|\nu_\im,\phi_\im\rangle\nonumber\\
&=& 2\pi\left[\langle \nu_\f,\phi_\f|\delta(p_\phi-\sqrt{\Theta})|\nu_\im,\phi_\im\rangle+\langle\nu_\f,\phi_\f|\delta(p_\phi+\sqrt{\Theta})|\nu_\im,\phi_\im\rangle\right]\,.\nonumber
\eea
Inserting decomposition (\ref{both}), one realizes that the first and second contributions involve, respectively, only positive- and negative-frequency states. Carrying out a $p_\phi$ integration explicitly, one arrives at the sum of definitions (\ref{newtonwig}) and (\ref{newtonwig2}) of the restricted Newton--Wigner functions, as announced.

We will see that the inclusion of the factor $2|p_{\phi}|$ into (\ref{hadaach}) changes the character of the inner product thus defined from relativistic to non-relativistic. In particular, we will show that the non-relativistic two-point function (\ref{newtonwig}) satisfies a different composition law than the Hadamard function (\ref{hadamard}). Therefore,
both the relativistic particle analogy and the composition laws satisfied by the different two-point functions are consistent with equation (\ref{rela}). However, starting from the two distinct expressions for the Newton--Wigner function $G_{{\rm NW}}^\pm$, one can proceed as in \cite{ach,achlong} and derive \emph{distinct} vertex expansions for the inner product in cosmology, which we will report below. This functional difference may come as a surprise \cite{ach,achlong}, but is what one could expect immediately when looking at the two expressions (\ref{hadaach}) and (\ref{newtonwig}), one involving the operator $\Theta$ in the exponent and the other its square root $H$. Here, equation (\ref{rela}) shows the existence of a \emph{unique} Newton--Wigner function in both the group averaging and deparametrized framework, thus settling the issue.


\subsection{Vertex expansions: Hadamard and Newton--Wigner functions}\label{vertexsec}

Evaluating the quantity (\ref{hadamard}) or (\ref{hadaach}), one observes \cite{achlong} that $\hat{C}$ consists of two pieces that only act on $\mathcal{H}_{{\rm kin}}^{{\rm g}}$ and $\mathcal{H}_{{\rm kin}}^{\phi}$ respectively, so that for equation  (\ref{hadamard}) the group-averaged inner product takes the form
\ben
G_{{\rm H}}(\nu_\f,\phi_\f;\nu_\im,\phi_\im)\equiv\int\limits_{-\infty}^{+\infty}\rmd\alpha\;A_{\rm H}(\Delta\phi;\alpha)A_\Theta(\nu_\f,\nu_\im;\alpha)\,,
\label{alphaint}
\een
where 
\ben
A_{\rm H}(\Delta\phi;\alpha)\equiv\int\limits_{-\infty}^{+\infty}\frac{\rmd p_{\phi}}{2\pi}\;\rme^{\im\alpha p_{\phi}^2}\rme^{\im p_{\phi}\Delta\phi}\,,\label{afi}
\een
while for equation  (\ref{hadaach}) we get \cite{ach}
\ben
G_{{\rm NW}}(\nu_\f,\phi_\f;\nu_\im,\phi_\im)\equiv\int\limits_{-\infty}^{+\infty}\rmd\alpha\; A_{\rm NW}(\Delta\phi;\alpha)A_\Theta(\nu_\f,\nu_\im;\alpha)\,,
\label{alphaint2}
\een
where
\ben
A_{\rm NW}(\Delta\phi;\alpha)\equiv\int\limits_{-\infty}^{+\infty}\frac{\rmd p_{\phi}}{2\pi}\;(2|p_{\phi}|)\rme^{\im\alpha p_{\phi}^2}\rme^{\im p_{\phi}\Delta\phi}\,.
\een
The factor $A_\Theta(\nu_\f,\nu_\im;\alpha)\equiv\langle \nu_\f|\rme^{-\im\alpha\Theta}|\nu_\im\rangle$ is common to both two-point functions, and their difference lies in the $p_\phi$-dependent part, that is, as it was reasonable to expect, in the different treatment of the `time' variable in the relativistic versus non-relativistic representation. 

In both cases, $A_\Theta(\nu_\f,\nu_\im;\alpha)$ is reorganized in the manner of Feynman by splitting up the `time' interval $\alpha$ into $N$ parts of equal length $\epsilon=\alpha/N$ and introducing decompositions of the identity on $\mathcal{H}_{{\rm kin}}^{{\rm g}}$,
\ben
\fl A_\Theta(\nu_\f,\nu_\im;\alpha) = \sum_{\bar{\nu}_{N-1},\ldots,\bar{\nu}_1} \langle\nu_\f|\rme^{-\im\epsilon\Theta}|\bar{\nu}_{N-1}\rangle \langle\bar{\nu}_{N-1}|\rme^{-\im\epsilon\Theta}|\bar{\nu}_{N-2}\rangle \ldots \langle\bar{\nu}_1|\rme^{-\im\epsilon\Theta}|\nu_\im\rangle\,.
\een
One then reorganizes the sum by characterising each possible history $(\nu_\im,\bar{\nu}_1,\ldots,\bar{\nu}_{N-2},\bar{\nu}_{N-1},\nu_\f)$ by the number of volume transitions $M$ that occur:
\bea
\fl A_\Theta(\nu_\f,\nu_\im;\alpha) & = & \sum_{M=0}^N \sum_{{\nu_{M-1},\ldots,\nu_1}\atop{\nu_m\neq \nu_{m+1}}} A_N(\nu_\f,\nu_{M-1},\ldots,\nu_1,\nu_\im;\alpha)
\\& \equiv & \sum_{M=0}^N \sum_{{\nu_{M-1},\ldots,\nu_1}\atop{\nu_m\neq \nu_{m+1}}} \sum_{N_M=M}^{N-1}\sum_{N_{M-1}=M-1}^{N_M-1}\ldots\sum_{N_1=1}^{N_2-1}(\langle \nu_\f |\rme^{-\im\epsilon\Theta}|\nu_\f\rangle)^{N-N_M-1} \nonumber
\\& &\times\langle \nu_\f |\rme^{-\im\epsilon\Theta}|\nu_{M-1}\rangle\ldots \langle\nu_1 |\rme^{-\im\epsilon\Theta}|\nu_\im\rangle(\langle \nu_\im |\rme^{-\im\epsilon\Theta}|\nu_\im\rangle)^{N_1-1}\,,
\label{expansion}
\eea
where $N_i$ denotes the step at which the $i$th transition occurs. While up to this point we have been using the LQC identity decomposition $\sum_{\nu}|\nu\rangle\langle\nu|$ in the derivation, one would obtain completely analogous expressions for Wheeler--DeWitt theory by replacing all summations $\sum_{\nu_i}$ by integrals $\int \rmd\nu_i$.

In the form (\ref{expansion}) the limit $N\rightarrow\infty$ can be defined rigorously;\footnote{This is, of course, the continuum limit of the discretization of the $\alpha$-interval, which is the counterpart of the continuum limit of the (proper) time interval for the relativistic particle. In the spin foam analogy, it would correspond, in the full theory, to the infinite-refinement limit of the triangulation on which the spin foam is based. Defining this limit in the full spin foam case, or in simplicial quantum gravity, is of course a highly non-trivial task. Equally non-trivial would be to show that this limit can equivalently written as a kind of sum over discretizations of the same topology (here, the line interval). In the (L)QC case this is much easier, and in fact the existence of a vertex expansion for the two-point function, defined in the same limit, is a proof of such equivalence, at least for this simple case.} the sums over $N_1,\ldots,N_M$ become integrals and one finally obtains
\ben
A_\Theta(\nu_\f,\nu_\im;\alpha) = \sum_{M=0}^{\infty} \sum_{{\nu_{M-1},\ldots,\nu_1}\atop{\nu_m\neq \nu_{m+1}}} A(\nu_\f,\nu_{M-1},\ldots,\nu_1,\nu_\im;\alpha)\,,
\een
where \cite{achlong} 
\bea
&&\fl A(\nu_\f,\nu_{M-1},\ldots,\nu_1,\nu_\im;\alpha)=\Theta_{\nu_\f\nu_{M-1}}\ldots\Theta_{\nu_2\nu_1}\Theta_{\nu_1 \nu_\im}\nonumber
\\&&\qquad\times\prod_{k=1}^p\frac{1}{(n_k-1)!}\left(\frac{\partial}{\partial \Theta_{w_k w_k}}\right)^{n_k-1} \sum_{m=1}^p\frac{\rme^{-\im\alpha\Theta_{w_m w_m}}}{\prod_{{j=1}\atop{j\neq m}}^p (\Theta_{w_m w_m}-\Theta_{w_j w_j})}\,.
\eea
Here,
\ben
\Theta_{\nu_k \nu_l}\equiv\langle\nu_k|\Theta|\nu_l\rangle
\een
are the matrix elements of $\Theta$, and in the second line the $p$ distinct values appearing in $(\nu_\im,\nu_1,\ldots,\nu_{M-1},\nu_\f)$ are denoted by $w_1,\ldots,w_p$ with multiplicities $n_i$ so that $n_1+\ldots+n_p=M+1$. It is this step that can only be done formally if the variable $\nu$ has a continuous range. Our analysis, which is rigorous for LQC, will be understood to hold for Wheeler--DeWitt theory in this formal sense.

In order to give the complete vertex expansion of the two-point functions, one has to insert the above expression in equations (\ref{alphaint}) and (\ref{alphaint2}), perform the integration over $\alpha$, and finally the one over $p_\phi$. 

The integration over $\alpha$ in (\ref{alphaint}) leads to delta distributions $\delta(p_{\phi}^2-\Theta_{w_m w_m})$ enforcing the Hamiltonian constraint. One is left with a straightforward integration over $p_{\phi}$, giving the final result
\bea
\fl G_{{\rm H}}(\nu_\f,\phi_\f;\nu_\im,\phi_\im)&=&\sum_{M=0}^{\infty} \sum_{{\nu_{M-1},\ldots,\nu_1}\atop{\nu_m\neq \nu_{m+1}}} \Theta_{\nu_\f\nu_{M-1}}\ldots\Theta_{\nu_2\nu_1}\Theta_{\nu_1 \nu_\im}\prod_{k=1}^p\frac{1}{(n_k-1)!}\nonumber
\\&&\times\left(\frac{\partial}{\partial \Theta_{w_k w_k}}\right)^{n_k-1} \sum_{m=1}^p\frac{1}{\sqrt{\Theta_{w_m w_m}}}\frac{\cos(\sqrt{\Theta_{w_m w_m}}\Delta\phi)}{\prod_{{j=1}\atop{j\neq m}}^p (\Theta_{w_m w_m}-\Theta_{w_j w_j})}\,.\label{hadamardresult}
\eea
The reader will observe the appearance of a cosine typical of the Hadamard two-point function.

Integrating over $\alpha$ in (\ref{alphaint2}), one notes that the delta distributions $\delta(p_{\phi}^2-\Theta_{w_m w_m})$ appearing under the $p_{\phi}$ integral are replaced by equation (\ref{delta}) with $\Theta\to \Theta_{w_m w_m}$, so that restricting to positive or negative $p_{\phi}$ corresponds to picking out positive- and negative-frequency solutions to the square root of the constraint (\ref{hamconst}) as one does for the Newton--Wigner function of the relativistic particle \cite{halliortiz}.

Integrating over both positive and negative $p_{\phi}$ one finds \cite{ach}
\bea
\fl G_{{\rm NW}}(\nu_\f,\phi_\f;\nu_\im,\phi_\im)&=&\sum_{M=0}^{\infty} \sum_{{\nu_{M-1},\ldots,\nu_1}\atop{\nu_m\neq \nu_{m+1}}} \Theta_{\nu_\f\nu_{M-1}}\ldots\Theta_{\nu_2\nu_1}\Theta_{\nu_1 \nu_\im}\nonumber
\\&&\times\prod_{k=1}^p\frac{1}{(n_k-1)!}\left(\frac{\partial}{\partial \Theta_{w_k w_k}}\right)^{n_k-1} \sum_{m=1}^p\frac{2\cos(\sqrt{\Theta_{w_m w_m}}\Delta\phi)}{\prod_{{j=1}\atop{j\neq m}}^p (\Theta_{w_m w_m}-\Theta_{w_j w_j})}\,.\label{achresult}
\eea
Once more, one recognizes the cosine factor typical of symmetrized (over frequency sectors) solutions of the Hamiltonian constraint, and the different functional dependence on the `energy' $H$ resulting from the non-relativistic representation of the canonical states.

Restricting to positive $p_{\phi}$ gives
\bea
\fl G_{{\rm NW}}^+(\nu_\f,\phi_\f;\nu_\im,\phi_\im)&=&\sum_{M=0}^{\infty} \sum_{{\nu_{M-1},\ldots,\nu_1}\atop{\nu_m\neq \nu_{m+1}}} \Theta_{\nu_\f\nu_{M-1}}\ldots\Theta_{\nu_2\nu_1}\Theta_{\nu_1 \nu_\im}\nonumber
\\&&\times\prod_{k=1}^p\frac{1}{(n_k-1)!}\left(\frac{\partial}{\partial \Theta_{w_k w_k}}\right)^{n_k-1} \sum_{m=1}^p\frac{\rme^{\im\sqrt{\Theta_{w_m w_m}}\Delta\phi}}{\prod_{{j=1}\atop{j\neq m}}^p (\Theta_{w_m w_m}-\Theta_{w_j w_j})}\,.\label{achresult2}
\eea
This procedure provides, in the simplified setting of loop quantum cosmology, an explicit spin-foam type expansion of the physical inner product $G_{{\rm H}}$ or $G_{{\rm NW}}$, thus strengthening (at least on a formal level and in this ultra-minimalistic model) the relation between canonical and covariant approaches.

From the expression for the deparametrized inner product (\ref{newtonwig}), it is immediate that it can be expressed in the form
\bea
\fl G_{{\rm NW}}^+(\nu_\f,\phi_\f;\nu_\im,\phi_\im) & = & \sum_{M=0}^{\infty} \sum_{{\nu_{M-1},\ldots,\nu_1}\atop{\nu_m\neq \nu_{m+1}}} A(\nu_\f,\nu_{M-1},\ldots,\nu_1,\nu_\im;\alpha)\Big|_{\alpha\leftrightarrow\Delta\phi,\Theta\leftrightarrow-H}\nonumber\\
& = & \sum_{M=0}^{\infty} \sum_{{\nu_{M-1},\ldots,\nu_1}\atop{\nu_m\neq \nu_{m+1}}} H_{\nu_\f\nu_{M-1}}\ldots H_{\nu_2\nu_1}H_{\nu_1 \nu_\im}\nonumber
\\&&\times\prod_{k=1}^p\frac{1}{(n_k-1)!}\left(\frac{\partial}{\partial H_{w_k w_k}}\right)^{n_k-1} \sum_{m=1}^p\frac{\rme^{\im H_{w_m w_m}\,\Delta\phi}}{\prod_{{j=1}\atop{j\neq m}}^p (H_{w_m w_m}-H_{w_j w_j})}\,,
\eea
where
\ben
H_{\nu_k \nu_l}\equiv\langle\nu_k|\sqrt{\Theta}|\nu_l\rangle\,.
\een
As noticed in \cite{ach}, group averaging and the deparametrized framework lead to distinct vertex expansions. However, as we have already discussed above, the two expansions only correspond to different perturbative expressions of the \emph{same} two-point function. This can be recognized working at the `non-perturbative' level, i.e., before operating the vertex expansion, or else (if one is able to) after this has been performed, by re-summing the whole series.


\subsection{Vertex expansions: Feynman propagator}

A most crucial issue in quantum gravity concerns different possible definitions for the transition amplitudes, and in particular the difference between two-point functions defining canonical inner products and \emph{causal} two-point functions defining, possibly, true transition amplitudes. However, discussions on the problem are usually constrained to take place on a formal level (see, e.g., \cite[chapter 1.3.1]{danielethesis} for an overview, \cite{Tei82, halliwellhartle} for the formal construction in quantum gravity path integrals, and \cite{causalBC,feynman,causal3dmatter} for tentative definitions of causal counterparts of spin foam models for quantum gravity). 

It may prove insightful to make such discussions explicit in a simplified model such as (loop) quantum cosmology, exploiting once more the analogy with the relativistic particle. In the latter case, one may define the Feynman Green's function in a way similar to equation (\ref{hadasumoverh}) by
\ben
\im G_{{\rm F}}(x'',t'';x',t')=\int\limits_{0}^{+\infty} \rmd T\;g(x'',t'';T|x',t';0)\,,
\een
so that in the gravitational context one only integrates over positive lapse. This suggests to define the correspondent quantity in quantum cosmology as
\ben
{\rm i}G_{{\rm F}}(\nu_\f,\phi_\f;\nu_\im,\phi_\im)\equiv\int\limits_{0}^{+\infty}\rmd\alpha\;A_{\rm H}(\Delta\phi;\alpha)A_\Theta(\nu_\f,\nu_\im;\alpha)\,,
\een
where $A_{\rm H}$ is given in equation (\ref{afi}) (no factor $2|p_{\phi}|$). Let us stress that the restriction in the Schwinger parameter $\alpha$, the (L)QC analogue of the proper time for the relativistic particle \cite{halliortiz} and of the lapse for quantum gravity \cite{Tei82}, is {\it not} a restriction in any time variable, e.g.~$\phi$, and it is thus fully consistent with (symmetry-reduced) background independence \cite{halliwellhartle}. As is usual for the Feynman propagator, one has to choose a contour in the complex $p_{\phi}$ plane since now the integrand has poles. We follow the causal `$+\im\varepsilon$' prescription:
\bea
\fl \im G_{{\rm F}}(\nu_\f,\phi_\f;\nu_\im,\phi_\im) = \sum_{M=0}^{\infty} \sum_{{\nu_{M-1},\ldots,\nu_1}\atop{\nu_m\neq \nu_{m+1}}} \Theta_{\nu_\f\nu_{M-1}}\ldots\Theta_{\nu_2\nu_1}\Theta_{\nu_1 \nu_\im}\prod_{k=1}^p\frac{1}{(n_k-1)!}\nonumber\\
\hspace{-1cm}\times\left(\frac{\partial}{\partial \Theta_{w_k w_k}}\right)^{n_k-1}\sum_{m=1}^p \int\limits_{-\infty}^{+\infty}\frac{\rmd p_{\phi}}{2\pi}\frac{\im}{p_{\phi}^2-\Theta_{w_m w_m}+\im\varepsilon}\frac{\rme^{\im p_{\phi}\Delta\phi}}{\prod_{{j=1}\atop{j\neq m}}^p (\Theta_{w_m w_m}-\Theta_{w_j w_j})}.\label{intermfeyn}
\eea
Performing the $p_{\phi}$ integration in the complex plane, we obtain
\bea
\fl \im G_{{\rm F}}(\nu_\f,\phi_\f;\nu_\im,\phi_\im) = \sum_{M=0}^{\infty} \sum_{{\nu_{M-1},\ldots,\nu_1}\atop{\nu_m\neq \nu_{m+1}}} \Theta_{\nu_\f\nu_{M-1}}\ldots\Theta_{\nu_2\nu_1}\Theta_{\nu_1 \nu_\im}\prod_{k=1}^p\frac{1}{(n_k-1)!}\nonumber
\\\hspace{-1cm} \times\left(\frac{\partial}{\partial \Theta_{w_k w_k}}\right)^{n_k-1} \sum_{m=1}^p \frac{\rme^{-\im\sqrt{\Theta_{w_m w_m}}\Delta\phi}\theta(\Delta\phi)+\rme^{\im\sqrt{\Theta_{w_m w_m}}\Delta\phi}\theta(-\Delta\phi)}{2\sqrt{\Theta_{w_m w_m}}\prod_{{j=1}\atop{j\neq m}}^p (\Theta_{w_m w_m}-\Theta_{w_j w_j})}\,.\label{feynmanresult}
\eea
Because of the open contour (or equivalently, because of the restriction to the positive semiaxis in the $\alpha$ integration) the two-point function $G_{{\rm F}}$ is {\it not} a solution of the Hamiltonian constraint equation; thus it does not define an inner product for the canonical theory. However, it is a proper Green's function (propagator) for the same equation, i.e.~it defines a transition amplitude taking into account the relative ordering in the `time' variables $\phi$ labelling the states (and thus defining a background-independent notion of `in' and `out').

We may also determine Wightman functions $G_{\pm}$ as in \cite{halliortiz} by choosing a closed contour around only one of the poles $p_{\phi}=\pm\sqrt{\Theta_{w_m w_m}}$. This obviously corresponds to restricting to positive- or negative-frequency sectors. One finds:
\bea
\fl G^\pm(\nu_\f,\phi_\f;\nu_\im,\phi_\im) & = & \sum_{M=0}^{\infty} \sum_{{\nu_{M-1},\ldots,\nu_1}\atop{\nu_m\neq \nu_{m+1}}} \Theta_{\nu_\f\nu_{M-1}}\ldots\Theta_{\nu_2\nu_1}\Theta_{\nu_1 \nu_\im}\prod_{k=1}^p\frac{1}{(n_k-1)!}\nonumber
\\& &\times \left(\frac{\partial}{\partial \Theta_{w_k w_k}}\right)^{n_k-1} \sum_{m=1}^p \frac{\rme^{\mp\im\sqrt{\Theta_{w_m w_m}}\Delta\phi}}{2\sqrt{\Theta_{w_m w_m}}\prod_{{j=1}\atop{j\neq m}}^p (\Theta_{w_m w_m}-\Theta_{w_j w_j})}.
\label{wightman}
\eea
The choice of a closed contour also implies that the two Wightman functions, like the Hadamard two-point function, satisfy the Hamiltonian constraint for (loop) quantum cosmology, and thus define appropriate physical inner products for the two sectors corresponding to positive- and negative-frequency, in the relativistic representation. 

One can easily show that the two-point functions we have defined satisfy the relations familiar from the relativistic particle case:
\bea
\fl G_{{\rm H}}(\nu_\f,\phi_\f;\nu_\im,\phi_\im)&=&G^+(\nu_\f,\phi_\f;\nu_\im,\phi_\im)+G^-(\nu_\f,\phi_\f;\nu_\im,\phi_\im)\,,\\
\fl \im G_{{\rm F}}(\nu_\f,\phi_\f;\nu_\im,\phi_\im)&=&G^+(\nu_\f,\phi_\f;\nu_\im,\phi_\im)\theta(\phi_\f-\phi_\im)+G^-(\nu_\f,\phi_\f;\nu_\im,\phi_\im)\theta(\phi_\im-\phi_\f)\,.
\eea
Finally, we can define the \emph{causal two-point function} by
\ben
\im G_{{\rm c}}(\nu_\f,\phi_\f;\nu_\im,\phi_\im)\equiv G^+(\nu_\f,\phi_\f;\nu_\im,\phi_\im)-G^-(\nu_\f,\phi_\f;\nu_\im,\phi_\im)\,,
\label{causal}
\een
noting that this gives the explicit expression
\bea
\fl G_{{\rm c}}(\nu_\f,\phi_\f;\nu_\im,\phi_\im) & = & -\sum_{M=0}^{\infty} \sum_{{\nu_{M-1},\ldots,\nu_1}\atop{\nu_m\neq \nu_{m+1}}} \Theta_{\nu_\f\nu_{M-1}}\ldots\Theta_{\nu_2\nu_1}\Theta_{\nu_1 \nu_\im}\prod_{k=1}^p\frac{1}{(n_k-1)!}\nonumber
\\& & \times\left(\frac{\partial}{\partial \Theta_{w_k w_k}}\right)^{n_k-1} \sum_{m=1}^p \frac{\sin\left(\sqrt{\Theta_{w_m w_m}}\Delta\phi\right)}{\sqrt{\Theta_{w_m w_m}}\prod_{{j=1}\atop{j\neq m}}^p (\Theta_{w_m w_m}-\Theta_{w_j w_j})}\,,\label{causalexplicit}
\eea
where the characteristic sine factor appears.

It is interesting to note that, once more in parallel with the relativistic particle case, one can obtain the above causal two-point function as the inner product between quantum states in the relativistic representation (\ref{both}) with completeness relation (\ref{dec}). This results in an indefinite inner product, contrary to the decomposition (\ref{dec2}) which produces the Hadamard function (\ref{alphaint}). In contrast to the latter, the causal two-point function $G_{\rm c}$ does not have a sum-over-histories representation \cite{halliortiz}.


\subsection{A local vertex expansion?}

A slightly different derivation for the inner product in loop quantum cosmology has been recently given in \cite{hrvw}, motivated by the following two shortcomings of the inner product (\ref{achresult}): Firstly, it was noted that the derivation of (\ref{achresult}) that we have sketched in section \ref{vertexsec} contains ambiguities related to a particular ordering for summations and integrations to be carried out. Since one is dealing with divergent quantities along the way (the integral $\int \rmd\alpha\;A(\nu_\f,\nu_{M-1},\ldots,\nu_1,\nu_\im;\alpha)$ is a sum of distributions and derivatives of distributions), there is no guarantee that the result will be independent of such choices. Secondly, the formal analogy of (\ref{achresult}) with a spinfoam amplitude lacks the locality familiar from spin foam models for full quantum gravity \cite{alex,danielethesis,baez}. These issues hold not only for equation (\ref{achresult}) but also for the other two-point functions.

Both issues are adressed by noting the following identity for two-point functions:
\ben
G_{{\rm H}}(\nu_\f,\phi_\f;\nu_\im,\phi_\im)=2 {\rm Re} \left[\im G_{{\rm F}}(\nu_\f,\phi_\f;\nu_\im,\phi_\im)\right]\,,
\een
which is true both for two-point functions for the relativistic particle given in \cite{halliortiz} and for (\ref{hadamardresult}) and (\ref{feynmanresult}).

This allows for a definition of the Hadamard function in terms of the Feynman propagator, which is now computed differently. Namely, starting from equation  (\ref{intermfeyn}), one performs the sum over $m$ in the second line before integrating, finding (this follows by taking expression (3.19) in \cite{hrvw}, substituting $\im\delta\rightarrow \im\varepsilon - p_{\phi}^2$ and taking the complex conjugate)
\bea
\fl \im G_{{\rm F}}^{{\rm HRVW}}(\nu_\f,\phi_\f;\nu_\im,\phi_\im) &=& \sum_{M=0}^{\infty} \sum_{{\nu_{M-1},\ldots,\nu_1}\atop{\nu_m\neq \nu_{m+1}}} \Theta_{\nu_\f\nu_{M-1}}\ldots\Theta_{\nu_2\nu_1}\Theta_{\nu_1 \nu_\im}\nonumber\\
&&\times\int\limits_{-\infty}^{+\infty}\frac{\rmd p_{\phi}}{2\pi}\frac{\im\, \rme^{\im p_{\phi}\Delta\phi}}{\prod_{k=0}^M (p_{\phi}^2-\Theta_{\nu_k \nu_k}+\im\varepsilon)}\,,
\label{hrvwresult}
\eea
where the products of differences of matrix elements in equations (\ref{hadamardresult}) and (\ref{feynmanresult}) have disappeared. The integrand now is {\em local}, inasmuch as it features a product over amplitudes of elements forming the particular `history of the Universe' one considers.

Since, in contrast to \cite{hrvw}, we have included the scalar field $\phi$, one is left with an integration over $p_{\phi}$ where one picks up residues at the poles $p_{\phi}=\pm\sqrt{\Theta_{w_i w_i}},\;i=1,\ldots,p$, which are of order $n_i$. It is not too difficult to see that the residue at $p_{\phi}=+\sqrt{\Theta_{w_m w_m}}$ is
\bea
\fl & & \frac{1}{(n_m -1)!}\frac{\partial^{n_m-1}}{\partial p_{\phi}^{n_m-1}}\left[\frac{1}{2\pi}\frac{\im\, \rme^{\im p_{\phi}\Delta\phi}(p_{\phi}-\sqrt{\Theta_{w_m w_m}})^{n_m}}{\prod_{k=0}^M (p_{\phi}^2-\Theta_{\nu_k \nu_k})}\right]\Bigg|_{p_{\phi}\rightarrow \sqrt{\Theta_{w_m w_m}}}\nonumber\\
\fl & = & \frac{1}{(n_m -1)!}\frac{\partial^{n_m-1}}{\partial p_{\phi}^{n_m-1}}\left[\frac{1}{2\pi}\frac{\im\, \rme^{\im p_{\phi}\Delta\phi}}{(p_{\phi}+\sqrt{\Theta_{w_m w_m}})^{n_m}\prod_{{j=1}\atop{j\neq m}}^p (p_{\phi}^2-\Theta_{w_j w_j})^{n_j}}\right]\Bigg|_{p_{\phi}\rightarrow \sqrt{\Theta_{w_m w_m}}}\nonumber\\
\fl & = & \prod_{{k=1}\atop{k\neq m}}^p \frac{1}{(n_k -1)!}\left(\frac{\partial}{\partial \Theta_{w_k w_k}}\right)^{n_k-1} \frac{1}{(n_m -1)!}\nonumber\\
\fl &&\times\frac{\partial^{n_m-1}}{\partial p_{\phi}^{n_m-1}}\left[\frac{1}{2\pi}\frac{\im\, \rme^{\im p_{\phi}\Delta\phi}}{(p_{\phi}+\sqrt{\Theta_{w_m w_m}})^{n_m}\prod_{{j=1}\atop{j\neq m}}^p (p_{\phi}^2-\Theta_{w_j w_j})}\right]\Bigg|_{p_{\phi}\rightarrow \sqrt{\Theta_{w_m w_m}}}\nonumber\\
\fl & = & \prod_{k=1}^p \frac{1}{(n_k -1)!}\left(\frac{\partial}{\partial \Theta_{w_k w_k}}\right)^{n_k-1} \left[\frac{1}{2\pi}\frac{\im\, \rme^{\im \sqrt{\Theta_{w_m w_m}}\Delta\phi}}{2\sqrt{\Theta_{w_m w_m}}\prod_{{j=1}\atop{j\neq m}}^p (\Theta_{w_m w_m}-\Theta_{w_j w_j})}\right]\,,
\eea
with a similar expression for $p_{\phi}=-\sqrt{\Theta_{w_m w_m}}$, so that one reproduces the `non-local' result (\ref{feynmanresult}). Therefore, while providing an interesting new viewpoint on the inner product for the purely gravitational sector, the findings of \cite{hrvw} do not modify our analysis. In the presence of free scalar matter, quantum cosmological two-point functions are always non-local, in the sense specified above.


\section{Composition laws}

Having computed the analogue for (loop) quantum cosmology of all two-point functions known for the relativistic particle, we are now in a position to verify the composition laws they satisfy and to interpret them in the cosmological setting. Such composition laws are in fact crucial for using these two-point functions as canonical inner products (or transition amplitudes). We will derive these laws using their sum-over-histories (or vertex expansion) formulation. It goes without saying that the results match the expectations coming from the particle analogy \cite{halliortiz}.


\subsection{Newton--Wigner function}

Let us start with the transition amplitude (\ref{newtonwig}) computed in the deparametrized framework. By inserting a resolution of the identity on the kinematical gravitational Hilbert space $\mathcal{H}_{{\rm kin}}^{{\rm g}}$, we immediately verify the composition law
\bea
G_{{\rm NW}}^+(\nu_\f,\phi_\f;\nu_\im,\phi_\im)&=&\langle \nu_\f|\rme^{\im H(\phi_\f-\phi_\im)}|\nu_\im\rangle\nonumber
\\& = & \sum_{\nu} \langle \nu_\f|\rme^{\im H(\phi_\f-\phi)}|\nu\rangle\langle\nu|\rme^{\im H(\phi-\phi_\im)}|\nu_\im\rangle\nonumber
\\& = & \sum_{\nu} G_{{\rm NW}}^+(\nu_\f,\phi_\f;\nu,\phi)\, G_{{\rm NW}}^+(\nu,\phi;\nu_\im,\phi_\im)
\eea
for any given intermediate $\phi$, which is the obvious counterpart of the usual non-relativistic composition law satisfied by the Newton--Wigner function,
\ben
G_{{\rm NW}}^+(x'',t'';x',t')=\int \rmd x\; G_{{\rm NW}}^+(x'',t'';x,t)G_{{\rm NW}}^+(x,t;x',t')\,.
\label{nonrelcomp}
\een
This rule incorporates the requirement that any path from $(x',t')$ to $(x'',t'')$ that only goes forward in time will pass through precisely one $x$ at a specified intermediate time $t'<t<t''$. Note that this identity is, as we have anticipated above, independent of the form of $H$ (or, equivalently, of $\Theta$).

Using the representation (\ref{alphaint2}) of the physical inner product in the timeless framework, a composition law of the form (\ref{nonrelcomp}) would amount to
\bea
\fl \int\limits_{-\infty}^{+\infty}\rmd\alpha\; A_{\rm NW}(\phi_\f-\phi_\im;\alpha)A_\Theta(\nu_\f,\nu_\im;\alpha)& \stackrel{?}{=}& \sum_{\nu}\int\limits_{-\infty}^{+\infty}\int\limits_{-\infty}^{+\infty}\rmd\alpha\;\rmd\alpha'\; A_{\rm NW}(\phi_\f-\phi;\alpha) A_{\rm NW}(\phi-\phi_\im;\alpha')\nonumber
\\& & \times A_\Theta(\nu_\f,\nu;\alpha)A_\Theta(\nu,\nu_\im;\alpha')\,.
\label{complaw?}
\eea
From the definition $A_\Theta(\nu_\f,\nu_\im;\alpha)\equiv\langle \nu_\f|\rme^{-\im\alpha\Theta}|\nu_\im\rangle$, it is again immediate that
\ben
\fl \sum_{\nu} A_\Theta(\nu_\f,\nu;\alpha)A_\Theta(\nu,\nu_\im;\alpha') = \sum_{\nu} \langle \nu_\f | \rme^{-\im\alpha\Theta}|\nu \rangle \langle \nu| \rme^{-\im\alpha'\Theta}|\nu_\im\rangle = A_\Theta(\nu_\f,\nu_\im;\alpha+\alpha')\,.
\een
Clearly, this composition rule extends to the case of Wheeler--DeWitt theory where $\nu$ has continuous range. The calculations in this section can be extended to the continuous case if one accepts the summation over $M$ to be of formal nature.

Further evaluating the right-hand side of (\ref{complaw?}), we have
\bea
\fl {\rm rhs\;of\;(\ref{complaw?})} & = & \int\limits_{-\infty}^{+\infty}\int\limits_{-\infty}^{+\infty}\rmd\alpha\;\rmd\alpha'\; A_{\rm NW}(\phi_\f-\phi;\alpha) A_{\rm NW}(\phi-\phi_\im;\alpha') A_\Theta(\nu_\f,\nu_\im;\alpha+\alpha')\nonumber
\\& = & \sum_{M=0}^{\infty} \sum_{{\nu_{M-1},\ldots,\nu_1}\atop{\nu_m\neq \nu_{m+1}}}\Theta_{\nu_\f\nu_{M-1}}\ldots\Theta_{\nu_2\nu_1}\Theta_{\nu_1 \nu_\im}\int \rmd\alpha\;\rmd\alpha'\;\frac{\rmd p_{\phi}}{2\pi}\frac{\rmd p'_{\phi}}{2\pi}\;(4|p_{\phi}p'_{\phi}|)\nonumber
\\&&\times \rme^{\im\alpha p_{\phi}^2}\rme^{\im\alpha'{p'_{\phi}}^2}\rme^{\im p_{\phi}(\phi_\f-\phi)}\rme^{\im p'_{\phi}(\phi-\phi_\im)}\prod_{k=1}^p\frac{1}{(n_k-1)!}\nonumber\\
&&\times\left(\frac{\partial}{\partial \Theta_{w_k w_k}}\right)^{n_k-1} \sum_{m=1}^p\frac{\rme^{-\im(\alpha+\alpha')\Theta_{w_m w_m}}}{\prod_{{j=1}\atop{j\neq m}}^p (\Theta_{w_m w_m}-\Theta_{w_j w_j})}\,,\label{int}
\eea
where the integration range of $p_{\phi}$ now depends on whether we restrict to positive frequency or not. 

Integrating over both positive and negative frequencies, equation (\ref{int}) reduces to
\bea
\fl &&\sum_{M=0}^{\infty} \sum_{{\nu_{M-1},\ldots,\nu_1}\atop{\nu_m\neq \nu_{m+1}}}\Theta_{\nu_\f\nu_{M-1}}\ldots\Theta_{\nu_2\nu_1}\Theta_{\nu_1 \nu_\im}\prod_{k=1}^p\frac{1}{(n_k-1)!}\nonumber
\\
\fl &&\times\left(\frac{\partial}{\partial \Theta_{w_k w_k}}\right)^{n_k-1} \sum_{m=1}^p\frac{4\cos[\sqrt{\Theta_{w_m w_m}}(\phi_\f-\phi)]\cos[\sqrt{\Theta_{w_m w_m}}(\phi-\phi_\im)]}{\prod_{{j=1}\atop{j\neq m}}^p (\Theta_{w_m w_m}-\Theta_{w_j w_j})}\,,
\eea
which is never equal to $G_{{\rm NW}}(\nu_\f,\phi_\f;\nu_\im,\phi_\im)$ because $\Theta$ is a positive operator. If we restrict to positive $p_{\phi}$, we will instead obtain
\bea
\fl \sum_{M=0}^{\infty} \sum_{{\nu_{M-1},\ldots,\nu_1}\atop{\nu_m\neq \nu_{m+1}}}&&\Theta_{\nu_\f\nu_{M-1}}\ldots\Theta_{\nu_2\nu_1}\Theta_{\nu_1 \nu_\im}\prod_{k=1}^p\frac{1}{(n_k-1)!}\nonumber\\
&&\times\left(\frac{\partial}{\partial \Theta_{w_k w_k}}\right)^{n_k-1} \sum_{m=1}^p\frac{\rme^{\im \sqrt{\Theta_{w_m w_m}}(\phi_\f-\phi)}\rme^{\im\sqrt{\Theta_{w_m w_m}}(\phi-\phi_\im)}}{\prod_{{j=1}\atop{j\neq m}}^p (\Theta_{w_m w_m}-\Theta_{w_j w_j})}\nonumber
\\&& = G_{{\rm NW}}^+(\nu_\f,\phi_\f;\nu_\im,\phi_\im)\,,
\eea
confirming that $ G_{{\rm NW}}^+$ as given in (\ref{achresult2}) does obey the non-relativistic composition law (\ref{nonrelcomp}). The same result will hold for the negative-frequency function $G_{{\rm NW}}^-$. On the other hand, the composition law is spoiled for $G_{{\rm NW}}$ due to the interference of positive- and negative-frequency sectors. 


\subsection{Hadamard function}

In the case of the relativistic particle, it is well known that two-point functions in the relativistic canonical representation, such as the Hadamard, Feynman and causal two-point functions, satisfy relativistic composition laws of the form \cite{halliortiz}
\ben
\tilde G(x''|x')=c\int_{\Sigma} \rmd\sigma^{\mu} G(x''|x)\stackrel{\leftrightarrow}{\partial}_{\mu}G(x|x')\,,
\een
where $G$ and $\tilde G$ can be different, typically $c=\pm 1$ or $c=\pm\im$, and $\Sigma$ is a spacelike surface in spacetime. In the present case, where the discretization of the $\nu$ coordinate breaks the $SO(1,1)$ isometry of a continuous (1+1)-dimensional spacetime, we take $\Sigma$ to be a surface $\phi=\const$, choose $G$, and try to verify the formula
\bea
\fl\tilde G(\nu_\f,\phi_\f;\nu_\im,\phi_\im)\stackrel{?}{=}c\sum_{\nu}\Big[G(\nu_\f,\phi_\f;\nu,\phi)\partial_{\phi}G(\nu,\phi;\nu_\im,\phi_\im)-G(\nu,\phi;\nu_\im,\phi_\im)\partial_{\phi}G(\nu_\f,\phi_\f;\nu,\phi)\Big]\nonumber\\
\label{formula}
\eea
for some $\tilde G$. First, taking $G=G_{{\rm H}}$, we have to show that the quantity
\bea
\fl &&\int\limits_{-\infty}^{+\infty}\int\limits_{-\infty}^{+\infty}\rmd\alpha\;\rmd\alpha'\left[A_{\rm H}(\phi_\f-\phi;\alpha)\partial_1 A_{\rm H}(\phi-\phi_\im;\alpha')+ A_{\rm H}(\phi-\phi_\im;\alpha)\partial_1 A_{\rm H}(\phi_\f-\phi;\alpha')\right]\nonumber\\
\fl &&\qquad\times A_\Theta(\nu_\f,\nu_\im;\alpha+\alpha')\,,
\label{complaw2?}
\eea
where $\partial_1$ is differentiation with respect to the first argument, is equal to $G_{{\rm c}}$ as in the point particle case, up to an overall factor. Substituting explicit expressions into (\ref{complaw2?}) we find
\bea
\fl {\rm (\ref{complaw2?})} & = & \sum_{M=0}^{\infty} \sum_{{\nu_{M-1},\ldots,\nu_1}\atop{\nu_m\neq \nu_{m+1}}}\int \rmd\alpha\;\rmd\alpha'\;\frac{\rmd p_{\phi}}{2\pi}\frac{\rmd p'_{\phi}}{2\pi}\;\rme^{\im\alpha p_{\phi}^2}\rme^{\im\alpha'{p'_{\phi}}^2}\rme^{\im p_{\phi}(\phi_\f-\phi)}\rme^{\im p'_{\phi}(\phi-\phi_\im)}\left(\im p_{\phi}+\im p'_{\phi}\right)\nonumber
\\
\fl &&\times\Theta_{\nu_\f\nu_{M-1}}\ldots\Theta_{\nu_2\nu_1}\Theta_{\nu_1 \nu_\im}\prod_{k=1}^p\frac{1}{(n_k-1)!}\left(\frac{\partial}{\partial \Theta_{w_k w_k}}\right)^{n_k-1} \sum_{m=1}^p\frac{\rme^{-\im(\alpha+\alpha')\Theta_{w_m w_m}}}{\prod_{{j=1}\atop{j\neq m}}^p (\Theta_{w_m w_m}-\Theta_{w_j w_j})}\nonumber
\\
\fl & = & -\sum_{M=0}^{\infty} \sum_{{\nu_{M-1},\ldots,\nu_1}\atop{\nu_m\neq \nu_{m+1}}}\Theta_{\nu_\f\nu_{M-1}}\ldots\Theta_{\nu_2\nu_1}\Theta_{\nu_1 \nu_\im}\prod_{k=1}^p\frac{1}{(n_k-1)!}\nonumber
\\
\fl &&\times\left(\frac{\partial}{\partial \Theta_{w_k w_k}}\right)^{n_k-1}\left[\sum_{m=1}^p\frac{\rme^{\im\sqrt{\Theta_{w_m w_m}}\Delta\phi}-\rme^{-\im\sqrt{\Theta_{w_m w_m}}\Delta\phi}}{2\im\sqrt{\Theta_{w_m w_m}}\prod_{{j=1}\atop{j\neq m}}^p (\Theta_{w_m w_m}-\Theta_{w_j w_j})}\right]\nonumber
\\
\fl & = & -\sum_{M=0}^{\infty} \sum_{{\nu_{M-1},\ldots,\nu_1}\atop{\nu_m\neq \nu_{m+1}}}\Theta_{\nu_\f\nu_{M-1}}\ldots\Theta_{\nu_2\nu_1}\Theta_{\nu_1 \nu_\im}\nonumber
\\
\fl &&\times\prod_{k=1}^p\frac{1}{(n_k-1)!}\left(\frac{\partial}{\partial \Theta_{w_k w_k}}\right)^{n_k-1}\sum_{m=1}^p\frac{\sin(\sqrt{\Theta_{w_m w_m}}\Delta\phi)}{\sqrt{\Theta_{w_m w_m}}\prod_{{j=1}\atop{j\neq m}}^p (\Theta_{w_m w_m}-\Theta_{w_j w_j})}\nonumber
\\
\fl & = & G_{{\rm c}}(\nu_\f,\phi_\f;\nu_\im,\phi_\im)\,,
\eea
where $G_{{\rm c}}$ is the causal two-point function (\ref{causalexplicit}). This is precisely the composition law one has for the Hadamard function of the relativistic particle \cite{halliortiz},
\ben
G_{{\rm c}}(x''|x')=\int_{\Sigma} \rmd\sigma^{\mu}\; G_{{\rm H}}(x''|x)\stackrel{\leftrightarrow}{\partial}_{\mu}G_{{\rm H}}(x|x')\,.
\een
The reason why the causal two-point function $G_{{\rm c}}$ and not $G_{{\rm H}}$ appears on the left-hand side is again interference between positive- and negative-frequency sectors.

We conclude that the two definitions used for the inner product in cosmology, the deparametrized definition (\ref{newtonwig}) and the timeless definition (\ref{hadamard}), indeed satisfy different composition laws, one being non-relativistic (after singling out positive or negative frequencies) and the other relativistic and involving the causal two-point function (\ref{causalexplicit}). 


\subsection{Feynman propagator}

For the Feynman Green's function, the composition law known for the relativistic particle is \cite{halliortiz}
\ben
\im G_{{\rm F}}(x''|x')=\im \int_{\Sigma} \rmd\sigma\; \im G_{{\rm F}}(x''|x)\stackrel{\leftrightarrow}{\partial}_{n}\im G_{{\rm F}}(x|x')\,,
\label{feynmcom}
\een
where $\stackrel{\leftrightarrow}{\partial}_{n}$ is a derivative taken along the normal in the direction of propagation. In quantum cosmology, its analogue should be (assuming $\phi_\im<\phi<\phi_\f$; one has an extra minus for $\phi_\f<\phi<\phi_\im$)
\bea
&&\fl \int\limits_{0}^{+\infty}d\alpha\;A_{\rm H}(\phi_\f-\phi_\im;\alpha)A_\Theta(\nu_\f,\nu_\im;\alpha) \stackrel{?}{=} \im\int\limits_{0}^{+\infty}\int\limits_{0}^{+\infty}\rmd\alpha\;\rmd\alpha'\;\left[A_{\rm H}(\phi_\f-\phi;\alpha)\partial_1 A_{\rm H}(\phi-\phi_\im;\alpha')\right.\nonumber
\\
\fl &&\qquad \left.+A_{\rm H}(\phi-\phi_\im;\alpha)\partial_1 A_{\rm H}(\phi_\f-\phi;\alpha')\right] A_\Theta(\nu_\f,\nu_\im;\alpha+\alpha')\,,
\label{complaw3?}
\eea
where $\partial_1$ means differentiation with respect to the first argument. Again, we can substitute explicit expressions into the right-hand side to obtain, using the Feynman contour in the complex plane:
\bea
\fl {\rm rhs\;of\;(\ref{complaw3?})} & = & \im\sum_{M=0}^{\infty} \sum_{{\nu_{M-1},\ldots,\nu_1}\atop{\nu_m\neq \nu_{m+1}}}\int\limits_0^{+\infty} \rmd\alpha\;\rmd\alpha'\int\limits_{-\infty}^{+\infty}\frac{\rmd p_{\phi}}{2\pi}\frac{\rmd p'_{\phi}}{2\pi}\;\rme^{\im\alpha p_{\phi}^2}\rme^{\im\alpha'{p'_{\phi}}^2}\rme^{\im p_{\phi}(\phi_\f-\phi)}\rme^{\im p'_{\phi}(\phi-\phi_\im)}\nonumber
\\
\fl &&\times\left(\im p_{\phi}+\im p'_{\phi}\right)\Theta_{\nu_\f\nu_{M-1}}\ldots\Theta_{\nu_2\nu_1}\Theta_{\nu_1 \nu_\im}\prod_{k=1}^p\frac{1}{(n_k-1)!}\nonumber\\
\fl &&\times\left(\frac{\partial}{\partial \Theta_{w_k w_k}}\right)^{n_k-1} \sum_{m=1}^p\frac{\rme^{-\im(\alpha+\alpha')\Theta_{w_m w_m}}}{\prod_{{j=1}\atop{j\neq m}}^p (\Theta_{w_m w_m}-\Theta_{w_j w_j})}\nonumber
\\ 
\fl & = & -\im\sum_{M=0}^{\infty} \sum_{{\nu_{M-1},\ldots,\nu_1}\atop{\nu_m\neq \nu_{m+1}}}\Theta_{\nu_\f\nu_{M-1}}\ldots\Theta_{\nu_2\nu_1}\Theta_{\nu_1 \nu_\im}\int\limits_{-\infty}^{+\infty}\frac{\rmd p_{\phi}}{2\pi}\frac{\rmd p'_{\phi}}{2\pi}\;\rme^{\im p_{\phi}(\phi_\f-\phi)}\rme^{\im p'_{\phi}(\phi-\phi_\im)}\nonumber
\\
\fl &&\times\left(\im p_{\phi}+\im p'_{\phi}\right)\prod_{k=1}^p\frac{1}{(n_k-1)!}\left(\frac{\partial}{\partial \Theta_{w_k w_k}}\right)^{n_k-1}\nonumber\\
\fl &&\times\left[\sum_{m=1}^p\frac{(p_{\phi}^2-\Theta_{w_m w_m}+\im\varepsilon)^{-1}({p'_{\phi}}^2-\Theta_{w_m w_m}+\im\varepsilon)^{-1}}{\prod_{{j=1}\atop{j\neq m}}^p (\Theta_{w_m w_m}-\Theta_{w_j w_j})}\right]\nonumber
\\ 
\fl & = & \frac{1}{2}\sum_{M=0}^{\infty} \sum_{{\nu_{M-1},\ldots,\nu_1}\atop{\nu_m\neq \nu_{m+1}}}\Theta_{\nu_\f\nu_{M-1}}\ldots\Theta_{\nu_2\nu_1}\Theta_{\nu_1 \nu_\im}\prod_{k=1}^p\frac{1}{(n_k-1)!}\left(\frac{\partial}{\partial \Theta_{w_k w_k}}\right)^{n_k-1}\nonumber
\\
\fl && \times\sum_{m=1}^p\frac{\rme^{-\im\sqrt{\Theta_{w_m w_m}}\Delta\phi}\theta(\phi_\f-\phi)\theta(\phi-\phi_\im)-\rme^{\im\sqrt{\Theta_{w_m w_m}}\Delta\phi}\theta(\phi_\im-\phi)\theta(\phi-\phi_\f)}{\sqrt{\Theta_{w_m w_m}}\prod_{{j=1}\atop{j\neq m}}^p (\Theta_{w_m w_m}-\Theta_{w_j w_j})}\nonumber
\\
\fl & = & \cases{\im G_{{\rm F}}(\nu_\f,\phi_\f;\nu_\im,\phi_\im)\,, & $\phi_\im<\phi<\phi_\f$\,, \cr -\im G_{{\rm F}}(\nu_\f,\phi_\f;\nu_\im,\phi_\im)\,, & $\phi_\f<\phi<\phi_\im$\,,}
\eea
in agreement with equation (\ref{feynmcom}). 

Choosing the closed contour used to define Wightman functions $G^{\pm}$ in (\ref{wightman}), one also confirms
\ben
G^{\pm}(x''|x')=\pm\im \int_{\Sigma} \rmd\sigma^{\mu}\; G^{\pm}(x''|x)\stackrel{\leftrightarrow}{\partial}_{\mu} G^{\pm}(x|x')\,.
\een
As we mentioned, the causal two-point function does not have a vertex expansion, but from its definition (\ref{causal}) one sees that it correctly obeys the same relativistic composition law.

To conclude, the quantities that we identified as being the analogue of Hadamard, Feynman, and Newton--Wigner two-point functions for quantum cosmology satisfy the same composition laws as one has for the relativistic particle in 1+1 dimensions, independently of the details of the operator $\Theta$.


\section{Discussion}

In this paper, we have catalogued the possible two-point functions for (loop) quantum cosmology, for a single homogeneous universe in the presence of a free massless scalar field, relying heavily on the analogy of this system with the free relativistic particle. For each of them we have constructed a covariant sum-over-histories representation, in the form of a vertex expansion in the sense of \cite{ach}. Finally, we have derived the composition laws that such two-point functions satisfy. Our results hold for homogeneous and isotropic LQC in a flat universe as well as for the traditional Wheeler--DeWitt quantum cosmology, although only in a more formal sense (due to the continuous nature of volume eigenvalues\footnote{Indeed, the same `vertex expansion' could be derived for the relativistic particle in flat space, again with a similar purely formal nature.}). Our results hopefully clarify (or at least provide a basis for clarifying) some issues that have been recently debated in the LQC literature. 

The features that characterize the different two-point functions are

\noindent (a) the sector of the canonical Hilbert space they refer to (positive- or negative-frequency, or both), 

\noindent (b) the relativistic or non-relativistic representation chosen for the quantum states, 

\noindent (c) their composition law, 

\noindent (d) their nature as canonical inner products or propagators (Green's functions).

If one restricts to a single sector of solutions, say positive-frequency, then both the relativistic (positive) Wightman function (\ref{wightman}) and the non-relativistic (positive) Newton--Wigner function (\ref{newtonwig}) provide a satisfactory and physically equivalent definition of the canonical inner product. In fact, they are positive definite and satisfy an appropriate composition law (of relativistic and non-relativistic type, respectively). The Feynman propagator (\ref{feynmanresult}), reduced in this case to the retarded propagator, provides instead the Green's function that propagates solutions of the constraint equation into other solutions. Notice that all these functions are complex-valued, and transform into their complex conjugate under switch of the ordering of their two arguments.

The choice between a relativistic and a non-relativistic representation, when available, is then purely a matter of practical convenience, and it depends on the physical issue one is set to address. 

The Newton--Wigner function $G_{\rm NW}^+$, for example, was studied in connection to the problem of localization for a quantum mechanical particle. Its associated inner product, in fact, allows one to consider wavefunctions whose modulus squared gives the probability to find a particle in a volume about a certain spatial point \cite{Rui}. This result, however, depends on the choice of a frame because the Newton--Wigner function is not Lorentz invariant.
In the case of $G_{\rm NW}^+$ in quantum cosmology, $\sum_{\nu\in I}|\Psi(\nu,\phi)|^2$ would be the probability to have a universe with a given matter configuration and with volume in a certain discrete range $I$. The frame choice for the relativistic particle now corresponds to the choice of the scalar field as an internal clock. This interpretation is apparent in the deparametrized framework. Because of the Lorentz invariance, on the other hand, the Wightman functions are more commonly used. In the LQC setting, no such invariance is apparent, though.

The choice of a deparametrized/non-relativistic representation, however, is only available in the presence of a degree of freedom that could play the role of a clock. This role is played by the massless scalar field we have also used, but clearly no similar degree of freedom is present (a) in the pure gravity case and, (b) for interacting (or even purely massive) matter fields,  in the presence of a non-trivial (non-monotonic) scalar potential with local extrema, if one wants to go beyond the perturbative regime. In the general case, therefore, the only available option  for a covariant definition of the canonical inner product is the (complex) Wightman function defined through a group averaging procedure.

If one does {\it not} (or cannot) restrict the attention to a single sector of solutions, and wants (or needs) to include both of them in the dynamics, then the situation is much trickier for canonical quantization. One simply does not have a two-point function that can be used as an adequate physical inner product. In fact, the only two-point function that defines a positive define inner product in the relativistic representation of the canonical theory, the Hadamard function (\ref{alphaint}), does not satisfy the appropriate composition law, and the same is true for the Newton--Wigner two-point function (\ref{newtonwig}) in the non-relativistic representation. On the other hand, the only two-point function that could, in principle,  define a canonical inner product (since it is a solution of the constraint equations) and that satisfies the needed composition law is the causal two-point function (\ref{causal}), which is not positive definite and thus not viable either. The Feynman propagator, which also satisfies a good composition law, is not a solution of the constraint equation, but a Green's function (true propagator) for the same. Of course, this is well known for the relativistic particle \cite{halliortiz}, and the same issues have been discussed at the formal level in the full quantum gravity case \cite{kuchar,isham}.

Clearly, the notion of `frequency' is tied to the presence of a scalar field and is not available in the pure gravity case, or generically whenever $p_\phi$ is not a Dirac observable.
The natural background-independent analogue of the notion of positive/negative frequencies, available also in the absence of matter fields as well as in the full (non symmetry-reduced) theory is that of spacetime orientation \cite{Tei82,halliwellhartle, causalBC,feynman,causal3dmatter,kuchar,isham}.

The question then becomes whether there is a consistent way of dealing with superpositions and interactions between degrees of freedom corresponding to the two orientation sectors, in the case in which one cannot specialize to one, in a canonical framework. The natural suggestion, also in the light of the particle analogy, is to move to a second-quantized formalism, in which both the number of degrees of freedom being excited at different times and their associated frequencies/orientations are allowed to fluctuate.  In fact, when considering relativistic quantum particles, one notices that conservation of the particle number is incompatible with Lorentz symmetry and the field theory formalism is naturally invoked. Then, the relativistic scalar product for the field is not only well defined but also uniquely so. Even though the cosmological or quantum gravity analogue of this symmetry argument is unclear to us, at present, this would suggest that also in the quantum cosmology case, as soon as more gravitational and matter degrees of freedom enter the picture (including anisotropies and inhomogeneities), and thus one moves closer to the full theory, the natural formalism is that of a field theory on (symmetry-reduced) superspace, a third quantization \cite{3rd}. This had been considered also in the full theory \cite{kuchar,isham}. This formalism would accommodate also another possible generalization of the framework considered in this paper, that is, the possibility of topology change. This entails the creation/annihilation of (possibly homogeneous and isotropic) universes, quanta of the superspace field naturally carrying both orientations (in analogy with the particle/antiparticle distinction). This framework would represent a rather important generalization of the quantum cosmology setting (possibly useful in addressing issues like the origin of the cosmological constant \cite{3rd,banks}). Any further discussion of this possible generalization would go beyond the scope of this paper. However, it is natural to draw the lesson of the relativistic many-particle system and ask ourselves the fate of the cosmological two-point functions in a multiverse picture. In this case, the single-particle/single-universe two-point functions do not need to satisfy any positive definiteness requirement (since they do not play the role of inner products), the non-relativistic representation of canonical states would simply not be available, and we would expect the Feynman propagator to be elected over the other choices by the requirements of causal propagation (positive proper time/lapse, ordering of arguments) and correct composition law. The causal two-point function would also play the useful role of the mean value of the commutator of (super-)field operators in the Fock vacuum. Thus, it would encode correlations on superspace, rather than being the propagator of particle/single-universe states. Work on such third quantized (loop) quantum cosmology is in progress \cite{giastefdan}.

To conclude, from our results in the symmetry-reduced setting, let us collect some lessons and suggestions for quantum gravity and for the group field theory (GFT)/spin foam approach in particular. Clearly, the step from symmetry-reduced models to full theory is huge, both technically and conceptually. Therefore, any insight we try to import from the former to the latter should be taken \emph{cum grano salis} and treated only as a suggestion to be explored further.

First, we have seen how important the choice of relativistic versus non-relativistic representation is for the quantum theory. In the covariant setting, the analogue of the spin foam constructions, this is reflected in the form of boundary states appearing in the sum-over-histories and in the composition law chosen for the same. We have pointed out that the non-relativistic formalism, although physically equivalent to the relativistic one, is only available in very special cases, and thus a relativistic representation is to be preferred. At the covariant level, this implies that one should try to construct relativistic transition amplitudes/inner products which compose by means of a relativistic composition law. The relativistic nature is characterized by the presence of `embedding information', i.e.~of the normal to the `canonical surface', in both the definition of the decomposition of the identity and the composition law for transition amplitudes/inner products. This is required for the correct implementation of covariance properties as well as for the correct composition of boundary data in the two-point functions. It is natural to conjecture that something similar happens in the full LQG theory and in the related spin foam models, and that the needed `embedding information' is to be represented by similar extrinsic data, in particular by the normal vector to the canonical hypersurface, in turn characterizing the extrinsic, i.e.~boost component of the gravitational connection. A generalization of LQG states to include such data has been proposed \cite{sergeietera, eteraprojected, sergei, eteramaite} and has proven crucial also in covariant spin foam constructions \cite{fk,eterasimone,danielesteffen,danielearistide}. So our results seem to support such generalization. 

Second, we have pointed out that the presence of matter fields always leads to a non-local vertex expansion. This non-locality is defined in contrast to the factorized form, in terms of single simplex contributions, that usual spin foam models take \cite{alex,baez}. In turn, this local form translates naturally in (or can be seen as a direct consequence of) the underlying GFT formulation of the same models \cite{iogft}. There are several attitudes one can take with respect to this fact: (a) One could expect a similar loss of locality in spin foam models coupled to matter, and thus the impossibility of a corresponding GFT formulation; (b) one can consider this non-locality as an artefact of the symmetry-reduced context (also considering that symmetry reduction is itself a rather non-local procedure) or of the specific type of matter considered. The second attitude is based also on the fact that the locality of spin foam models and GFTs has a clear spacetime interpretation, in that spin foam vertices or their dual simplices represent regions or events in a discrete spacetime, while the vertex expansion in cosmology lacks such interpretation exactly because of the prior symmetry reduction. The available examples of spin foam models and GFTs for gravity coupled to matter  \cite{danielehendryk,simoneYM,PR1,winston,danielejimmy,richard} seem to support the second possibility, although the issue should be considered as still open. Note also that non-locality is indeed due to the presence of a matter field, and not to its use as internal time, since two-point functions are non-local both in the deparametrized and timeless frameworks. Perturbative inhomogeneities often appear in the dynamics in a form similar to matter fields (e.g.~\cite{MMW}), so that also they might lead to non-local expressions; however, this is speculative as the two-point function classification is clear only in purely homogeneous models.

Third, we have stressed that the crucial difference in constructing and understanding two-point functions in quantum cosmology is whether one restricts to positive or negative-frequency solutions or keeps both, and that the analogue of this frequency condition in full quantum gravity would be the spacetime orientation. In spin foam models and GFTs, one has to specify orientation data for all the discrete elements of the spacetime substratum used in the quantum theory, i.e.~edges and faces of the spin foam 2-complex, or simplices and sub-simplices of the dual triangulation. Moreover, opposite orientation data for all these elements have to be included for consistency of composition properties. This seems to suggest that no analogue of the frequency restriction is available in the full theory (as in the case of general inhomogeneous or anisotropic cosmologies, or for generic matter fields). Of course, one can still consider amplitudes that either are symmetric under spacetime orientation (parity or time) reversal (in analogy with the Hadamard function) or transform into their own complex conjugate under reversal (in analogy with Feynman or causal two-point functions). Known spin foam models \cite{alex}, based on  constraining quantum BF theory, seem to be naturally of the symmetric type. On the other hand, constructions of non-symmetric (complex) and causally restricted transition amplitudes\footnote{That is, amplitudes obtained by restricting the range of proper time/lapse integration.} for LQG states have been considered in the spin foam framework \cite{causalBC,feynman,causal3dmatter}, and could be useful to investigate this issue further. 
In fact, quantum BF theory itself gives rise to symmetric (real) transition amplitudes that can be consistently interpreted as covariant definitions of the inner product of topological gravity. In contrast, causally restricted models are associated with true propagators or transition amplitudes that have an inbuilt notion of in- and out-states consistent with background independence (inasmuch as this notion is encoded in orientation data only and no time variable is needed). The situation in the full theory seems thus to be at odds with the results in the quantum cosmology setting: while the latter would suggest that a symmetric transition amplitude that is both positive definite and includes both sectors of solutions would fail to compose properly, this seems not to be the case for BF spin foam models. On the other hand, one could be tempted to attribute this situation to the triviality, in terms of dynamical degrees of freedom, of BF theory itself. However, a similar orientation independence is present in the Barrett--Crane model which, if coming from a BF-type construction, is not a topological model. Similarly puzzling is the situation, concerning this orientation business, from the GFT perspective. GFTs are third quantized frameworks \cite{iogft} where the fundamental field is defined on simplicial superspace, the space of geometries for a single simplex (a building block of a discrete space), and in which creation/annihilation of simplices and topology change are naturally described. Hence, one would expect that the corresponding Feynman amplitudes be naturally complex and interpreted as built out of propagators, rather than solutions, to the canonical constraints. This is, again, {\it not} the case for current GFT models, that generate the symmetric spin foam amplitudes we have just discussed. On the other hand, it would be the case for any effective GFT dynamics around non-trivial background configurations \cite{eterawinston,danieleeteraflorian,GFTcoherent}, which naturally translates into non-trivial (and singular) free propagators, analogous to the Feynman two-point function of ordinary quantum field theory. 

Much remains to be understood about these issues, and the simplified quantum cosmology setting could be a very useful ground for clarifying them, leading to a better development and understanding of a full quantum theory of gravity.


\section*{Acknowledgements}
This work is funded by the A Von Humboldt Stiftung, through a Sofja Kovalevskaja Prize, which is gratefully acknowledged. SG also acknowledges support from EPSRC, Trinity College, Cambridge, and the Max Planck Society.


\end{document}